\documentclass[aps,pra,reprint,groupedaddress]{revtex4-1}

\usepackage{color}
\usepackage{float}
\usepackage{graphicx}
\usepackage[position=top, caption=false]{subfig}
\usepackage{mathtools}
\usepackage[utf8]{inputenc}
\usepackage{gensymb}
\usepackage{hyperref}

\begin{document}


\title{Linear Pulse Compansion using Co-propagating Space-Time Modulation}


\author{Nima Chamanara}
\author{Christophe Caloz}
\affiliation{Polytechnique Montr\'{e}al, Montr\'{e}al, Qu\'{e}bec H3T 1J4, Canada.}


\date{\today}

\begin{abstract}
This paper presents a pulse compansion, i.e. compression or expansion, technique based on co-propagating space-time modulation. An engineered asymmetric space-time modulated medium, co-propagating with a pulse compands the pulse continuously and at a constant rate. The space-time medium locally modifies the velocity of different sections of the pulse in order to shape the pulse as it propagates. There is no theoretical limit on the compansion factor with the proposed system. Moreover, it can be designed to transform the pulse shape and its modulation linearly, without any distortion. Therefore the proposed technique can be used for up or down-conversion of modulated pulses, with extreme conversion ratios. The presented compansion technique is linear with respect to the input wave and therefore can be used to perform compansion or frequency conversion on multiple pulses simultaneously. 
\end{abstract}


\maketitle


\section{Introduction}

Pulse compansion, i.e. compression or expansion, is ubiquitous in physics and engineering. Pulse compression has important implications. It corresponds to a better resolution in spectroscopy, imaging, radar and sonar~\cite{saleh1991fundamentals, jepsen2011terahertz, grischkowsky1990far, skolnik1970radar, mittleman1999recent}. Moreover, it is associated with a higher throughput in  communication systems~\cite{proakis1994communication, lathi1998modern} or a higher peak power in pulsed lasers~\cite{gu2010high}. Pulse expansion is important in applications where excessive pulse energy could damage the system. In such cases high energy pulses can be expanded to decrease their instantaneous energy, end then be compressed back again either by the communication channel or the receiver~\cite{skolnik1970radar, martinez1987design}. 

\begin{figure}[ht!]
	\centering
	\subfloat{
		\label{fig:banner-conventional-method}
		\includegraphics[page=1]{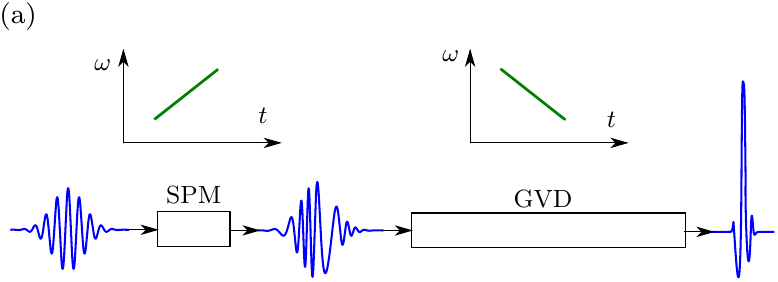}
	}
	\vfill
	\subfloat{
		\label{fig:banner-new-method}
		\includegraphics[page=2]{figs.pdf}
	}
	\caption{Conventional vs ST compansion. (a) Conventional compression. The wave modifies its own spectral content through self phase modulation (SPM) in a nonlinear medium, getting chirped in the process. The chirped pulse is then passed through a dispersive medium with opposite group velocity dersion (GVD) to compress the pulse. (b) Co-propagating ST pulse compression. A co-propagating ST medium (represented in red) compresses the pulse. The pulse keeps its shape and content during the compression without any distortion. The structure is linear with respect to the input waves, therefore multiple pulses (represented in blue) are compressed or expanded simultaneously.}
	\label{fig:banner}
\end{figure}

Conventional methods for pulse compression involve spectral broadening~\cite{boyd2003nonlinear, nisoli1997compression, seidel2016all} through self phase modulation in nonlinear Kerr media~\cite{boyd2003nonlinear, alfano1970observation, stolen1978self}, producing a chirped pulse. The chirp is then passed through a dispersive medium to get dechirped and compress the pulse~\cite{saleh1991fundamentals, boyd2003nonlinear, shank1982compression, finot2016design, brabec2000intense}. This process is schematically represented in Fig.~\ref{fig:banner-conventional-method}. However, the self-phase-modulation-induced chirp, as will be shown later, is irregular, and fundamentally, the resulting chirp can not be perfectly un-chirped. This fundamental limitation which is closely related to the underlying \emph{self} effect, leads to distorted spectra and consequently to imperfect pulses, limiting the compression factor. 

Pulse expansion is normally achieved by passing the pulse through a dispersive structure. As different frequency components propagate with different velocities, the resulting pulse envelop is broadened~\cite{martinez1987design, caloz2013analog}. However, the pulse gets chirped which may be undesirable in some applications.

This paper presents a technique for compressing and expanding pulses based on co-propagating space-time modulation. Space-time (ST) modulated media are materials whose parameters are controlled spatio-temporally. Such media are endowed with interesting properties. They don't conserve energy, as electromagnetic energy is pumped in or out of the medium through the modulation~\cite{cassedy1963dispersion, cassedy1967dispersion} and therefore can be employed in wave amplification~\cite{cullen1958travelling, cullen1960theory}. They naturally break Lorentz reciprocity~\cite{yu2009complete, sounas2013giant} and have found applications in nonreciprocal devices such as circulators and isolators~\cite{chamanara2017optical, estep2014magnetic, taravati2017nonreciprocal}. And finally they exhibit unusual properties such as pure mixing \cite{chamanara2017electromagnetic} unusual forward-forward coupling~\cite{chamanara2018unusual} and peculiar time refraction~\cite{akbarzadeh2018inverse}.

We use space-time modulation to locally control velocity of different parts of the pulse and mold its shape. As the underlying control mechanism does \emph{not} involve self effects, it provides a better degree of control over the compansion process. Therefore compansion can be achieved with a higher precision, where the envelop and contents of the pulse are transformed in a linear fashion, avoiding any distortion. Moreover, ST media are linear with respect to input waves, and therefore several pulses can be companded simultaneously. The proposed technique can be used for frequency up-down conversion of modulated electromagnetic pulses as well, without any fundamental limit on the compansion or frequency conversion ratios. Finally as compansion, as well as frequency mixing, are done based on the same principle, the proposed system has a good potential to be compact and programmable. Figure~\ref{fig:banner} compares co-propagating ST compression with conventional compression based on self effects.

The organization of the paper is as follows. Section~\ref{sec:principle} presents the compansion principle. Section~\ref{sec:spectral-transformation} discusses spectral transformations and applications in frequency conversion. Section~\ref{sec:spm-comparision} compared the proposed technique to the self phase modulation. Section~\ref{sec:results} provides simulation results. Section~\ref{sec:realization} presents a technique for experimental realization of the proposed STM compansion. And finally conclusions are presented in Sec.~\ref{sec:conclusion}.

\section{Compansion Principle} \label{sec:principle}

\begin{figure}[ht!]
	\centering
	\subfloat{ \label{fig:conc-pulse-medium-mov-frame-compress}
		\includegraphics[page=3]{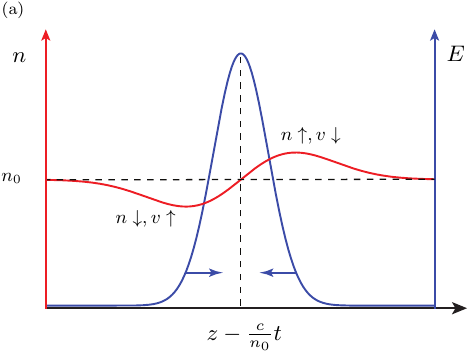}}
	\vfill
	\subfloat{  \label{fig:conc-pulse-medium-mov-frame-expand}
		\includegraphics[page=4]{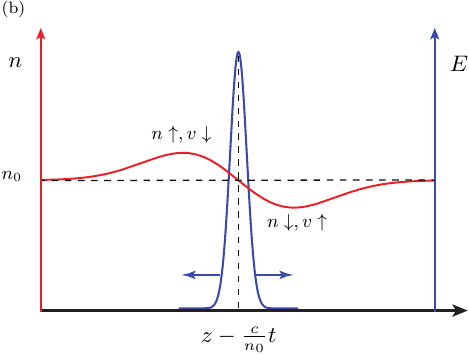}}
	\caption{Principle of compansion based on co-propagating ST modulation. Blue and red curves represent electric field and ST refractive index, respectively. (a) Compression. The leading part of the pulse sees a higher refractive index compared to its center and decelerates while its trailing part sees a lower refractive index and accelerates towards the center, leading to compression. (b) Expansion. The leading and trailing parts of the pulse accelerate and decelerate away, respectively, from its center, leading to expansion.}
	\label{fig:conc-pulse-medium-mov-frame}
\end{figure}

Consider an unmodulated Gaussian pulse, as shown in blue in Fig.~\ref{fig:conc-pulse-medium-mov-frame-compress}, propagating in a linear \emph{nondispersive} medium. We assume further that the medium is controlled spatio-temporally with an \emph{asymmetric} profile, shown in red in the same figure, co-propagating with the pulse. 

\begin{subequations}	\label{eq:n-zt}
	\begin{align}
	n\left(\zeta \right) &= n_0 + \Delta n\left(\zeta\right) \\
	\zeta &= z - \frac{c}{n_0}t
	\end{align}
\end{subequations}

\noindent
Note that the he horizontal axis represent the moving parameter $\zeta$. As the leading part of the pulse sees a higher refracting index compared to its center, it propagates with a slower velocity compared to its center. Therefore, the leading part of the pulse will compress towards its center. However, as it compresses towards the center, its velocity approaches the velocity of the center of the pulse. Therefore, the leading part of the pulse never crosses the center and will continuously compress. Inversely, the trailing part of the pulse sees a lower refractive index compared to its center and therefore accelerates towards the center. Therefore the pulse compresses brom both sides, symmetrically, and in a continuous rate.

The process can be reversed by simply flipping the sign of the modulation $\Delta n$, as shown in Fig.~\ref{fig:conc-pulse-medium-mov-frame-expand}. In this case the leading part of the pulse sees a lower refractive index and accelerates away from its center, while its trailing part sees a higher refractive index and decelerates away from the center. The overall effect is an expanding pulse.

\begin{figure}[ht!]
	\centering
	\subfloat{ \label{fig:conc-pulse-medium-st-diagram-compress}
		\includegraphics[page=5]{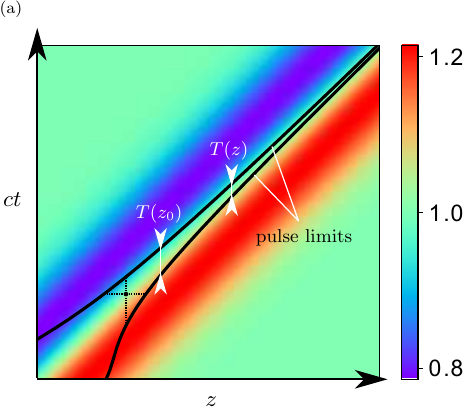}}
	\vfill
	\subfloat{ \label{fig:conc-pulse-medium-st-diagram-expand}
		\includegraphics[page=6]{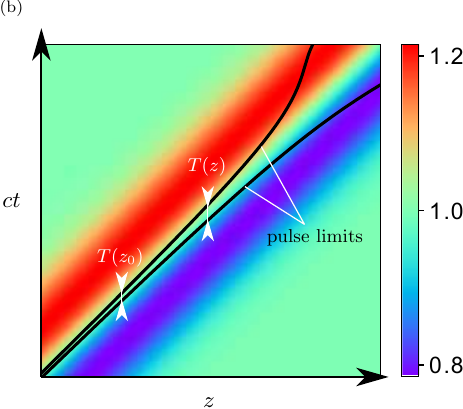}}
	\caption{Evolution of compansion in a space-time diagram. The color-plot represents the propagating ST modulated refractive index. (a) Compression. The leading edge of the pulse slows down and its trailing edge speeds up, leading to compression. (b) Expansion. The leading edge speeds up and the trailing edge slows down leading to expansion.}
	\label{fig:conc-pulse-medium-st-diagram}
\end{figure}

Space-time evolution of the ST compansion process is represented schematically in Fig.~\ref{fig:conc-pulse-medium-st-diagram}, where the vertical axis represents temporal evolution of plotted parameters. The 2D color-plot represents evolution of the ST medium, propagating along the  $+z$ direction with a constant velocity. The solid black lines represent the limits of the pulse (its extension). The ST medium and the pulse are co-propagating with the same velocity. Figure~\ref{fig:conc-pulse-medium-st-diagram-compress} represents pulse compression. The leading edge of the pulse sees a higher refractive index, and decelerates towards the pulse center, while the trailing edge sees a lower refractive index, and accelerates towards the pulse center, leading to compression. Similarly, Fig.~\ref{fig:conc-pulse-medium-st-diagram-expand} schematizes evolution of the expansion process.

\begin{figure}[ht!]
	\centering
	\subfloat{ \label{fig:conc-pulse-space-vs-time-T-cut}
		\includegraphics[page=7]{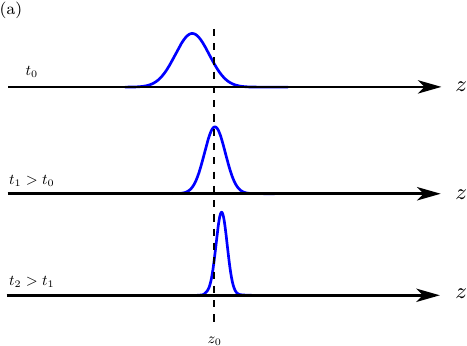}}
	\vfill
	\subfloat{ \label{fig:conc-pulse-space-vs-time-Z-cut}
		\includegraphics[page=8]{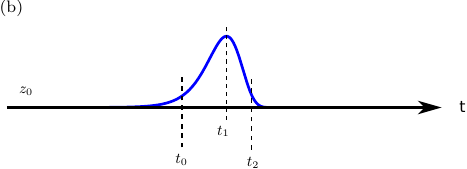}}
	\caption{Pulse evolution in space versus its evolution in time. (a) The pulse compresses but at any given instant it is symmetric. A fixed point in space measures pulses with different compression ration as the pulse passes. (b) The point $z_0$ measures three different pulses at instants $t_0, t_1$ and $t_2$, and therefore the pulse appears asymmetric in time.}
	\label{fig:conc-pulse-space-vs-time}
\end{figure}

Note that at any given time the pulse edges are symmetric with respect to its center located on the line $z = \frac{c}{n_0}t$, i.e. for a horizontal cut, represented by the dotted line in Fig.~\ref{fig:conc-pulse-medium-mov-frame-compress}, the right and left edges of the pulse have the same distance from its center point, represented by the black circle in Fig.~\ref{fig:conc-pulse-medium-mov-frame-compress}. However, the pulse edges are not symmetric with respect to a vertical cut. In other words the pulse is symmetric in space but not in time. Figure~\ref{fig:conc-pulse-space-vs-time} explains this effect. As the modulation is symmetric, the pulse remains symmetric at all times, as shown in Fig.~\ref{fig:conc-pulse-space-vs-time-T-cut}. However, a fixed point in space sees a pulse that is gradually compressing and therefore measures different pulses at different times. Therefore, the pulse appears asymmetric in time as shown in Fig.~\ref{fig:conc-pulse-space-vs-time-Z-cut}.

\section{Spectral Transformation} \label{sec:spectral-transformation}

\begin{figure}[ht!]
	\centering
	\subfloat{ \label{fig:conc-modulated-st-compress}
		\includegraphics[page=9]{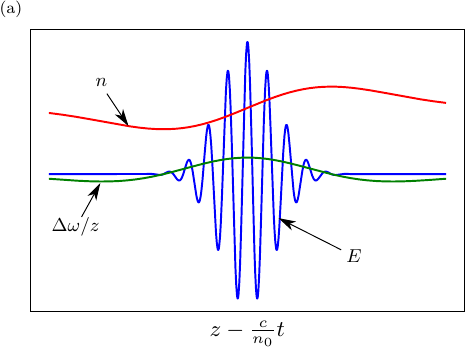}}
	\vfill
	\subfloat{ \label{fig:conc-modulated-st-expand}
		\includegraphics[page=10]{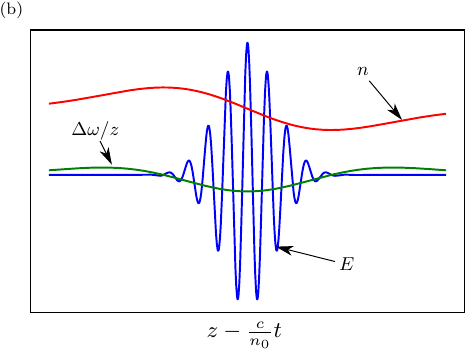}}
	\caption{Frequency conversion per unit length of the ST medium for a modulated Gaussian pulse. (a) Frequency up-conversion in a co-propagating ST medium with positive slope with respect to $\zeta$. (b) Frequency down-conversion in a co-propagating ST medium with negative slope.}
	\label{fig:conc-modulated-st}
\end{figure}

This section computes the change in frequency content of a modulated pulse as it co-propagates with a ST medium. Consider a modulated Gaussian pulse, shown in blue in Fig.~\ref{fig:conc-modulated-st-compress}, co-propagating with a ST medium shown in red. The space-time varying phase at any given ST point reads,

\begin{equation}	\label{eq:phase-zt}
\phi\left(z, t \right) = \omega_0 t - k z = \omega_0 t - k_0 n\left(z, t \right) z,
\end{equation}

\noindent
where $k_0=\omega_0/c$, and the ST refractive index $n(z, t)$ is given in \eqref{eq:n-zt}. Instantaneous frequency at any given ST point is then given by

\begin{equation}	\label{eq:omega-zt}
\omega\left(z, t \right) = \frac{\partial}{\partial t} \phi\left(z, t \right) = \omega_0 - k_0 z \frac{\partial}{\partial t} n\left(z, t \right).
\end{equation}

\noindent
Substituting \eqref{eq:n-zt} in \eqref{eq:omega-zt}, reduces \eqref{eq:omega-zt} to

\begin{equation}	\label{eq:omega-zeta}
\omega\left(z, t\right) = \omega_0 + \frac{2 \pi}{\lambda_0} \left[\frac{\partial}{\partial \zeta} \Delta n\left(\zeta \right)\right]z,
\end{equation}

\noindent
where $\lambda_0$ is the free space wavelength at $\omega_0$. Therefore the change in frequency per unit length is given by 

\begin{equation}	\label{eq:delta-omega-zeta}
\frac{\Delta\omega}{z} \left(\zeta\right) = \frac{\omega\left(z, t\right) - \omega_0}{z} = \frac{2 \pi}{\lambda_0} \left[\frac{\partial}{\partial \zeta} \Delta n\left(\zeta \right)\right],
\end{equation}

\noindent
which is plotted in Fig.~\ref{fig:conc-modulated-st} in green. A ST refractive index with a positive slope in $\zeta$, as in Fig.~\ref{fig:conc-modulated-st-compress}, up-shifts the frequency as the pulse co-propagates with the ST medium, whereas a negative slope, as in Fig.~\ref{fig:conc-modulated-st-expand} down-shifts the frequency. 

Depending on the slope of the ST medium different parts of the pulse may up-shift with different rates, as shown in the green curve in Fig.~\ref{fig:conc-modulated-st}. To get a uniform frequency shift, corresponding to a uniform compansion factor, the ST medium must maintain a relatively constant slope on the co-propagating pulse.

\section{Comparison with Self Phase Modulation} \label{sec:spm-comparision}

\begin{figure}[ht!]
	\centering
	\subfloat{	\label{fig:conc-modulated-spm-Enw}
		\includegraphics[page=11]{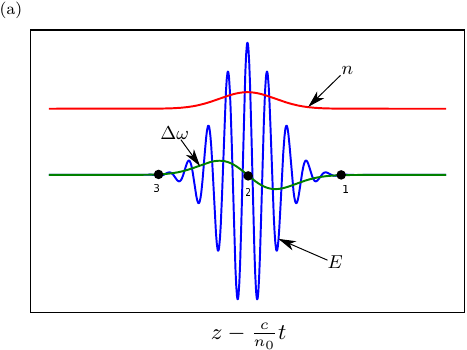}}
	\vfill
	\subfloat{	\label{fig:conc-modulated-spm-GD}
		\includegraphics[page=12]{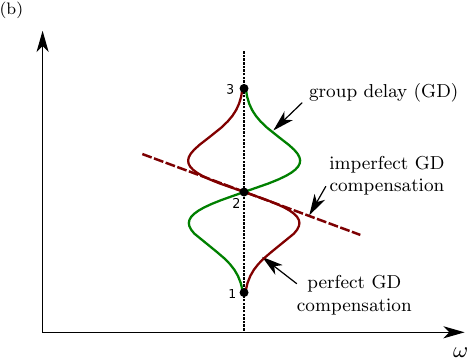}}
	\caption{Self phase modulation (SPM) in a nonlinear Kerr medium. (a) A modulated Gaussian pulse propagating in a Kerr medium. The red curve represents the effective refractive index, proportional to the intensity of the pulse. Frequency change per unit length of the medium is proportional to the green curve. (b) The green curve represents group delay dispersion. Solid brown curve represents GD dispersion of a dispersive medium that perfectly compresses the chirp. Such a multivalued GD dispersion is not realizable. The dashed brown curved represents a realizable GD dispersion. }
	\label{fig:conc-modulated-spm}
\end{figure}

Self phase modulation is a nonlinear effect where a strong pulse propagating in a nonlinear Kerr medium modulates its own phase, and therefore modifying its own spectral content. The nonlinear refractive index in a Kerr medium is proportional to the intensity of the pulse. Therefore for a modulated Gaussian pulse, shown in blue in Fig.~\ref{fig:conc-modulated-spm-Enw}, propagating in a Kerr medium, the effective refractive index will be a symmetric Gaussian shown in red in the same figure. Therefore, following \eqref{eq:delta-omega-zeta}, its frequency change per unit length will be asymmetric as shown in green in the same figure. As a result the pulse frequency content downshifts in its leading part and upshifts in its trailing part, i.e. the pulse gets chirped. The chirp is then unwound by passing the pulse through a medium with opposite gtoup delay dispersion, leading to compression. 

However, it is evident from Fig.~\ref{fig:conc-modulated-spm-Enw}, that the resulting chirp is irregular, as points 1, 2 and 3 remain at the same frequency. Therefore the center of the pulse is up-chirped while its edges are down-chirped. The corresponding group delay experienced by different frequency components is plotted in green in Fig.~\ref{fig:conc-modulated-spm-GD}. Note that points 1, 2 and 3 carry the same modulation frequency, while point 1 arrives earliest at the output and point 3 latest. Perfect chirp compensation in such a scenario requires a medium with group delay dispersion plotted in solid brown. Such a group delay dispersion is multi-valued and can not be realized. In practice such a group delay dispersion is approximated by a medium exhibiting a GD dispersion shown in the dashed brown curve. Therefore, the group delay is compensated only partially, at the central region of the pulse.

It should also be noted that SPM is a nonlinear effect, and therefore it does not support the superposition principle. Whereas, ST modulation is a linear effect. Therefore, it is possible to compress or expand multiple pulses simultaneously in the proposed compansion scheme based on ST modulation, however, in nonlinear compression schemes different signals will affect each other in an undesirable fashion.

\section{Results}	\label{sec:results}

This section presents full wave simulation results based on the finite difference time domain (FDTD) technique~\cite{taflove2005computational, chamanara2018simultaneous, vahabzadeh2018generalized}. The incident wave is a Gaussian pulse with the following waveform

\begin{equation}	\label{eq:fdtd-incident-pulse}
E\left(\zeta \right) = E_0 e^{-\left(\frac{\zeta}{\xi_E}\right)^2} \cos\left(k_0 \zeta\right),
\end{equation}

\noindent
launched at $z=z_0$, where $\omega_0 = k_0/c$ is the modulation frequency and $\zeta$ is the moving parameter given in \eqref{eq:n-zt}. Such a pulse has spatial and temporal widths proportional to $\xi_E$ and $\xi_E n_0/c$, respectively. Note that an unmodulated Gaussian pulse corresponds to $k_0 = 0$.

The co-propagating ST medium takes the following Gaussian derivative form

\begin{equation}	\label{eq:fdtd-st-medium}
n\left(\zeta \right) = n_0 \pm M \zeta e^{-\left(\frac{\zeta}{\xi_n}\right)^2},
\end{equation}

\noindent
where $M$ represents modulation depth, $\xi_n$ spatial extent of the modulation, and the $+$ and $-$ signs correspond to compression and expansion, respectively.

\subsection{Compression}

\begin{figure}[ht!]
	\centering
	\subfloat{ \label{fig:fdtd-compress-start-end}
		\includegraphics[page=13]{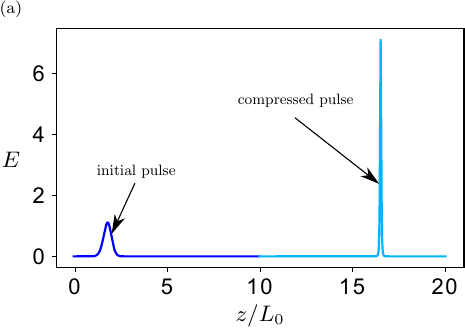}}
	\vfill
	\subfloat{ \label{fig:fdtd-compress-st}
		\includegraphics[page=14]{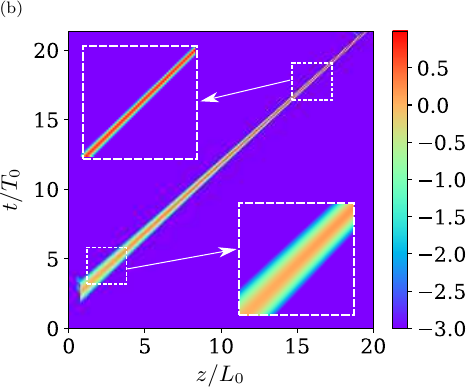}}
	\caption{Simulation of co-propagating ST pulse compression. (a) Initial and compressed pulses. (b) Space-time evolution of the pulse as it co-propagates with the ST medium. The color-plot represents the pulse amplitude in logarithmic scale. $L_0$ and $T_0$ are spatial and temporal widths of the initial pulse, respectively.}
	\label{fig:fdtd-pulse-compress}
\end{figure}

Figure~\ref{fig:fdtd-pulse-compress} shows FDTD simulation results for an unmodulated Gaussian pulse getting compressed through a co-propagating ST medium with modulation depth $M=0.01$. Figure~\ref{fig:fdtd-compress-start-end} shows the pulse in the beginning, and at the end of the ST medium. The pulse is compressed by a factor of 10 after propagating a distance 15 times its initial spatial width.

Figure~\ref{fig:fdtd-compress-st} shows the evolution of the pulse as it passes through the ST medium. The pulse gets progressively narrower at a linear rate as it propagates. Note that theoretically there is no limit on the level of compression. As long as the pulse and the ST medium are aligned and co-propagate with the same velocity the pulse continues to get narrower.

\subsection{Expansion}

\begin{figure}[ht!]
	\centering
	\subfloat{ \label{fig:fdtd-expand-start-end}
		\includegraphics[page=15]{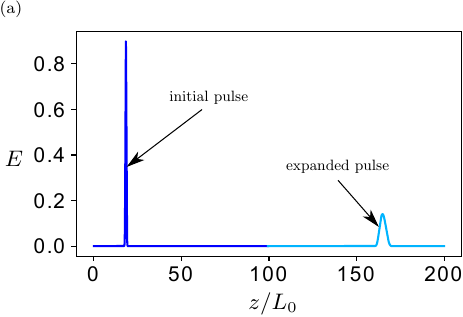}}
	\vfill
	\subfloat{ \label{fig:fdtd-expand-st}
		\includegraphics[page=16]{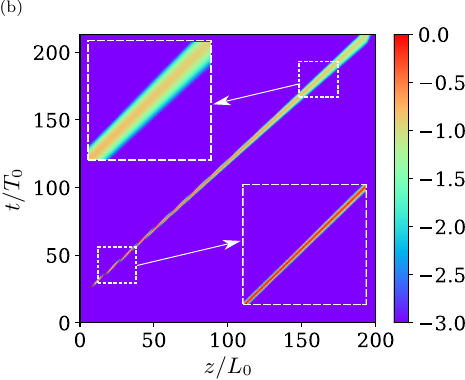}}
	\caption{Simulation of co-propagating ST pulse expansion. (a) Initial and expanded pulses. (b) ST evolution of the pulse as it co-propagates with the ST medium. The color plot represents the pulse amplitude in logarithmic scale. $L_0$ and $T_0$ are spatial and temporal widths of the initial pulse, respectively.}
	\label{fig:fdtd-pulse-expand}
\end{figure}

Figure~\ref{fig:fdtd-pulse-expand} shows simulation results for the expander ST medium. Figure~\ref{fig:fdtd-expand-start-end} shows the pulse at the beginning and at the end of the ST medium, for the modulation depth $M=0.01$. The pulse has expanded by a factor of 10 after propagating a distance 150 times its initial width.

Figure~\ref{fig:fdtd-expand-st} shows ST evolution of the pulse as it co-propagates with the ST medium. The pulse width broadens linearly as it propagates through the ST medium. Note that the pulse expands until its edges reaches the points corresponding to the maximum/minimum of the ST medium, beyond which it will start to distort. Thus, for higher expansion ratios a medium with a wider ST profile must be used.

\subsection{Frequency Conversion}

\begin{figure}[ht!]
	\centering
	\subfloat{ \label{fig:fdtd-modpulse-compress-start-end}
		\includegraphics[page=17]{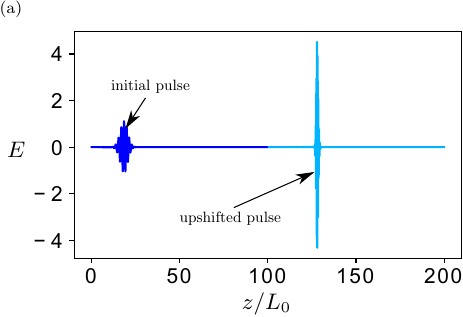}}
	\vfill
	\subfloat{ \label{fig:fdtd-modpulse-compress-spectrum}
		\includegraphics[page=18]{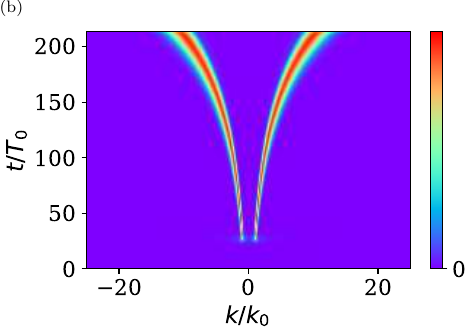}}
	\caption{Frequency up-conversion. $T_0$ is the temporal width of the initial pulse, and $k_0$ is its wave number. (a) The initial and up-shifted pulses. (b) Evolution of the spectrum of the pulse as it propagates through the ST medium. The color-plot is in linear scale.}
	\label{fig:fdtd-modpulse-compress}
\end{figure}

Figure~\ref{fig:fdtd-modpulse-compress} shows simulation results for a modulated pulse. The pulse gets up-shifted as it co-propagating with a compressive ST medium. Figure~\ref{fig:fdtd-modpulse-compress-start-end} shows the initial and up-shifted pulses, for the modulation depth $M=0.01$.

Figure~\ref{fig:fdtd-modpulse-compress-spectrum} represents space-time evolution of the spatial spectrum of the pulse as it propagates through the ST medium. A horizontal cut represents the Fourier transform of the pulse at that instant. At $t = 200 T_0$ the pulse is up-shifted by a factor 10.

\section{Realization}	\label{sec:realization}

Space-time modulation may be realized in various waves including nonlinearity~\cite{saleh1991fundamentals, boyd2003nonlinear, wang2013optical}, electro-optics~\cite{davis2014lasers, gottlieb1983electro, lin2015electro, ge2009electro}, acoustic or elastic waves~\cite{trainiti2016non}, acousto-optics~\cite{korpel1996acousto}, electrically tunable materials such as graphene~\cite{novoselov2005two, ferrari2006raman, geim2010rise, neto2009electronic, chamanara2013non, chamanara2012optically, chamanara2013terahertz, chamanara2016graphene, chamanara2015fundamentals} or some metal oxides~\cite{kinsey2015epsilon, ferrera2017dynamic} or electrically tunable circuit elements such as varactors~\cite{chamanara2017optical}.

\begin{figure}[ht!]
	\centering
	\subfloat{	\label{fig:realization-signals}
		\includegraphics[page=19]{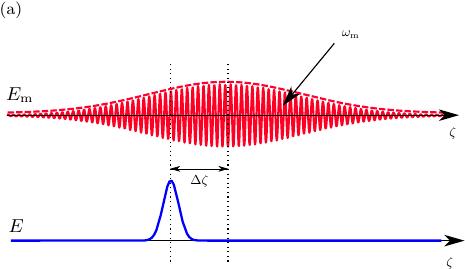}}
	\vspace{1cm}
	\vfill
	\subfloat{	\label{fig:realization-system}
		\includegraphics[page=20]{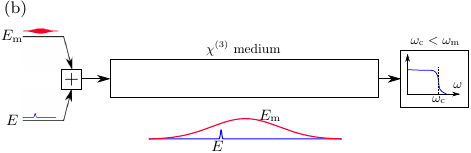}}
	\caption{Realization of ST pulse compression in a nonlinear Kerr medium. (a) $E$ represent the pulse to be compressed, and $E_m$ is a much stronger modulated pulse that excites the nonlinearity. The amplitudes are not to scale. $E$ is delayed with respect to $E_m$.  (b) $E_m$ and $E$ are injected in a nonlinear Kerr medium, and $E_m$ is filtered out at the output.  $E_m$ produces an effective ST medium that has a positive slope with respect to the moving parameter $\zeta$ and co-propagates with $E$.}
	\label{fig:realization}
\end{figure}

Figure~\ref{fig:realization} presents a method for realizing the proposed compansion techniques through cross phase modulation (XPM) in a nonlinear Kerr medium. Figure~\ref{fig:realization-signals} shows the waves that are injected in the system, and Fig.~\ref{fig:realization-system} presents the proposed compressive system. $E_\text{m}$ is a strong wave that excites the nonlinearity and produces an effective co-propagating ST medium, and $E$ is the pulse to be compressed. Note that $E$ is delayed with respect to $E_m$, such that $E$ is aligned at a point where $E_m$'s envelop, in red, has a positive slope. To later filter out $E_\text{m}$, it is modulated at a frequency outside the spectrum of $E$.

As shown in Fig.~\ref{fig:realization-signals}, $E_m$ and $E$ are injected in a nonlinear Kerr medium. The strong wave $E_m$ produces an effective ST medium proportional to its intensity, in red. Therefore, $E$ sees an effective co-propagating ST medium with a positive slope. This positive slope corresponds to the center of the Fig.~\ref{fig:conc-pulse-medium-mov-frame-compress}. As presented in Fig.~\ref{fig:conc-pulse-medium-mov-frame} a positive slope with respect to $\zeta$ leads to compression, while a negative slope leads to expansion. Therefore, $E$ is expected to compress as it co-propagates with $E_m$ inside the Kerr medium. $E_m$ is filter out at the end leaving behind the compressed wave $E$.

Note that the process outlined in Fig.~\ref{fig:realization} has some notable differences to conventional XPM. In conventional XPM, both waves affecting each other are usually of the same magnitude order, whereas in Fig.~\ref{fig:realization} $E_m$ is much stronger than $E$. Moreover, in contrast to Fig.~\ref{fig:realization}, in conventional systems XPM is usually a parasitic phenomenon with undesirable effects.

\section{Conclusions}	\label{sec:conclusion}
A pulse compansion technique based on co-propagating ST modulation has been presented. An engineered asymmetric ST modulated medium, co-propagating with a pulse has been shown to compresses or expands the pulse continuously and at a constant rate. The proposed technique is not subject to any theoretical limit, and can be used to compand electromagnetic pulses to extremely high ratios. The same technique may also be used for up or down-conversion of modulated pulses, with extreme conversion factors. The technique is linear and therefore can be used to perform compansion or frequency conversion on multiple signals simultaneously. Theoretical predictions have been validated with full wave simulation results.

\bibliography{ReferenceList2_abbr}

\end{document}